# Regional Frequency Constrained Dispatch Method Considering Spatial-joint Stochastic Disturbances and Contingencies

Nian Liu, *Member, IEEE*, Yubing Chen, *Student Member, IEEE*, Kai Jiang, *Member, IEEE*, Jiahao Liu, *Member, IEEE,* Cheng Wang, *Senior Member, IEEE*, and Tianshu Bi, *Fellow, IEEE*

*Abstract* — **The increasing penetration of renewable energy challenges frequency stability due to high variability and declining inertia. Traditional frequency-security-constrained dispatch methods fail to capture regional frequency heterogeneity and spatially correlated stochastic disturbances, resulting in inaccurate frequency security enforcement. To address this, a regional frequency-constrained dispatch method is proposed, considering the spatial-joint stochastic disturbances and contingencies. Firstly, a multi-regional frequency response model is constructed, incorporating the Vine-Copula-based characterization of regional stochastic disturbances and regional frequency support. Then, a progressive latent-bottleneck physics-informed neural network is applied to characterize differential frequency nadir terms and regional frequency support integral terms via an encoder–decoder architecture, which enables optimization compatible expressions of system frequency dynamics. Finally, a day-ahead dispatch model is developed in which regional stochastic frequency constraints are embedded using a CVaR-based formulation. A case study on the IEEE 118-bus system shows that the proposed method outperforms the unified center-of-inertia–embedded approach in mitigating regional frequency violations, with the regional frequency nadir and RoCoF improved by 32.76% and 2.06%, respectively.**

## NOMENCLATURE

### Indexes and Sets

| | |
|---|---|
| $t/T$ | Index/set of dispatch time slot. |
| $\tau$ | Index of transient time slot. |
| $i/\Omega_i$ | Index/set of all units. |
| $z/Z$ | Index/set of all regions. |
| $e/E_m$ | Index/set of the edge in the $m$-th vine tree. |
| $ep/E^{ep}$ | Index/number of epochs. |

### Variables

| | |
|---|---|
| $H_z^{sys}$ | Regional equivalent inertia constant. |
| $\Delta f$ | Frequency deviation value (Hz). |
| $P_i^{PFR}$ | The first frequency response power (MW). |
| $\Delta P_z^{export}$ | Deviation regional transmission power exported from region $z$ (MW). |
| $P_z^{export}/\tilde{P}_z^{export}$ | Regional transmission power under steady-state/after disturbance (MW). |
| $V_{z_i}$ | Equivalent voltage amplitude (V). |
| $\delta_{z_i}/\tilde{\delta}_{z_i}$ | Equivalent voltage phase angle under steady-state/after disturbance (°). |
| $\theta_z$ | Electrical angle of region z (rad). |
| $\omega_z/\omega_s$ | Rotational speed of region z/under steady-state (rad/s). |
| $P_{i,t}^H$ | Inertia response power (MW). |
| $P_{i,t}^{PFR}$ | Primary frequency response power (MW). |
| $U_{i,t}^G$ | Unit commitment decisions for synchronous generators. |
| $S_i^{G/WT/ESS}$ | Maximum available generation capacity of synchronous generators/WTs/ESSs (MW). |
| $\text{RoCoF}_z$ | Rate of change of frequency (Hz/s). |
| $f_z^{nadir}$ | Frequency nadir value (Hz). |
| $R_{i,t}^{G/WT/ESS}$ | Primary frequency reserve capacity (MW). |
| $\gamma_z$ | The residual losses of the neural network. |
| $L_z^{physics}$ | Physics-informed regularization term. |
| $C_{i,t}^{gen/UC/R}$ | Generation costs/unit start or off costs/primary frequency reserve costs ($). |
| $P_{i,t}^{G/re/dis/ch}$ | Power output of thermal generators/ renewable energy unit/ESSs (MW). |
| $x_{i,t}^{G_on/G_off}$ | Start-up/ shut-down indicator. |

### Parameters and constants

| | |
|---|---|
| $\Delta P_{z,t}$ | Equivalent disturbance (MW). |
| $\Delta\tilde{P}_{z,t}^s$ | Stochastic disturbances bias induced by the forecast errors in renewable generation (MW). |
| $P_z^0$ | Regional contingencies value (MW). |
| $h_z$ | Bandwidth for the kernel density estimation (KDE) function. |
| $\rho_e$ | Correlation coefficient between the pair-copulas of edge $e$. |
| $f_0$ | Steady-state frequency (Hz). |
| $D_z$ | Load damping coefficient (MW/Hz). |
| $X_{z_i z_j}$ | Reactance between region $z_i$ and region $z_j$ (Ω). |
| $H_i^{G/WT/ESS}$ | Inertia response coefficient of synchronous generators/WTs/ESSs (MW·s/Hz). |

This work is supported by the National Natural Science Foundation of China under Grants (U24B6008), (Corresponding author: Kai Jiang, kai_jiang@ncepu.edu.cn)

N. Liu, Y. Chen, K. Jiang, C. Wang, and T. Bi are with the State Key Laboratory of Alternate Electrical Power System with Renewable Energy Sources, North China Electric Power University, Beijing 102206, China (e-mail: nianliu@ncepu.edu.cn;chenyubing713@ncepu.edu.cn;kai_jiang@ncepu.edu.cn; chengwang@ncepu.edu.cn; tsbi@ncepu.edu.cn ).

J. Liu is with the School of Electrical & Electronic Engineering, University College Dublin, Belfield, Ireland (jiahao.liu@ucd.ie).

| | |
|---|---|
| $T_i^{G/WT/ESS}$ | Primary frequency response time constants of synchronous generators/WTs/ESSs (s). |
| $K_i^{G/WT/ESS}$ | The droop control response coefficient of generators/WTs/ESSs (MW/Hz). |
| $\boldsymbol{W}_{(\cdot)}/\boldsymbol{b}_{(\cdot)}$ | Weight matrix and bias vector of the *n*-th layer of the neural network. |
| $\sigma_{PINN}$ | Trainable parameters of the PLB-PINN. |
| $\alpha$ | Weights for the network residual losses. |
| $\overline{RoCoF}$ / $\overline{f^{nadir}}$ | Security threshold of the RoCoF value (Hz/s)/frequency nadir value (Hz). |
| $\varepsilon^{RoCoF}$ / $\varepsilon^{nadir}$ | Maximum exceeding risk level of the RoCoF and frequency nadir. |
| $a_i/b_i/c_i$ | Synchronous generators' generation cost coefficient ($/MW). |
| $\lambda_i^{start}/\lambda_i^{off}$ | Synchronous generators' startup/shutdown cost coefficient ($/Time). |
| $\lambda_i^{R_th/WT/ESS}$ | Synchronous generators'/WTs'/ESSs/ first primary reserve capacity cost coefficient ($/MW). |
| $P_{i,t}^{load}$ | Load demand (MW). |
| $\underline{P}_{z_i z_j}/\overline{P}_{z_i z_j}$ | Minimum/maximum transmission section capacity (MW). |
| $\underline{P_i^G}/\overline{P_i^G}$ | Minimum/maximum thermal generators' output limitation (MW). |
| $R_i^{UP}/R_i^{DN}$ | Ramp-up/ramp-down limitations (MW). |
| $t_i^{on}/t_i^{off}$ | Minimum start-up/shut-down durations (min). |
| $\eta_{i,t}^{dis}/\eta_{i,t}^{ch}$ | Discharging and charging efficiencies. |
| $\overline{P_i^{ESS}}$ | Maximum power output of ESS (MW). |
| $\underline{E_{i,t}^{ESS}}/\overline{E_{i,t}^{ESS}}$ | Lower/upper and bounds of the ESSs capacity (MWh). |
| $\overline{P_i^{wt}}/\overline{P_i^{pv}}$ | Maximum available power from WTs/PVs (MW). |

**Functions and operators**

| | |
|---|---|
| $\mu_\tau$ | Unit step function. |
| $f(\cdot)$ | Probability density function (PDF). |
| $F(\cdot)$ | Cumulative distribution function (CDF). |
| $K(\cdot)$ | The KDE function. |
| $\Phi(\cdot)$ | CDF of two-dimensional normal distribution. |
| $c(\cdot)$ | Pair-copula density function. |
| $C(\cdot)$ | Gaussian copula distribution function. |
| ReLU($\cdot$) | Rectified Linear Unit. |

# I. Introduction

With the rapid growth of renewable energy and power-electronic-interfaced resources, power systems are characterized by insufficient regulation and frequency supporting ability [1]. Under such conditions, dispatch decisions that are optimal from an energy-balance perspective may lead to inadequate inertia or poor frequency support [2], which may cause unacceptable frequency violations [3]. Thereby, the adequate inertia and frequency reserve capacities should be considered in the economic operation of system dispatch [4].

In frequency-security-constrained dispatch, the spatial heterogeneity of frequency support has become increasingly significant [5-6]. With the increasing penetration of renewable energy, frequency support resources are unevenly distributed across regions, leading to pronounced differences in regional inertia levels, primary frequency response capabilities, and frequency resilience. Regions with high renewable penetration usually exhibit lower inertia and weaker inherent frequency support capability [7], whereas regions dominated by synchronous generators can provide stronger inertial and primary frequency support, albeit with relatively slower response characteristics [8]. After a disturbance, these regional differences may result in non-uniform frequency dynamics, where frequency security indicator in weak regions can violate their limits even though the system-level frequency remains within the acceptable range. Moreover, inter-regional frequency support is constrained by the available transmission capacity of tie-lines. When tie-line capacity is limited, the frequency support capability in one region cannot be fully transferred to another frequency-weak region. Therefore, heterogeneous regional frequency characteristics should be explicitly considered in the dispatch model.

Recently, extensive efforts have been devoted to the frequency security-embedded dispatching problems [9-10]. To model the correlations between dispatch strategies and frequency response process, the resources-oriented frequency support model is constructed [11-17]. The frequency security is satisfied through diverse supporting resources, including virtual inertia [11], droop control [12], and fast frequency response [13] from converter-interfaced generation [14], wind turbines [15-16], and energy storage systems [17]. However, the complex integral representation of nodal frequency oscillations and differential formulation of the frequency swing equation make it challenging to directly embed the resulting nonlinear frequency constraints into the dispatch problem. To tackle challenges above, the center-of-inertia (COI) is applied based on the unified system-level frequency representation, which avoids the frequency integral terms. Also, the differential expression of constraints is linearized through the generic average system frequency model [18], the Bernstein polynomials method [19], and convolutional neural networks [20]. However, the above methods cannot be directly extended to regional frequency security-embedded dispatch problems with heterogeneous frequency support demands and capabilities. The COI representation ignores the spatial distribution of frequency support resources across areas, potentially resulting in localized frequency violations even when system-level frequency security appears to be satisfied.

Beyond spatial heterogeneity, the increasing uncertainty of renewable energy output further exacerbates the challenge of the frequency security-embedded dispatch problem. Although renewable energy output uncertainty has been extensively investigated in day-ahead dispatch problems [21], it also plays a critical role in the frequency security-embedded problem. Large forecast errors can introduce disturbances comparable to the traditional contingencies and cannot be neglected [22]. For example, in the Great Britain transmission system operated by National Grid ESO, short-term wind power forecast errors can reach 1 GW, which is a magnitude comparable to major generation contingencies on the order of 1-1.5 GW [23].

Along this line, existing studies incorporate stochastic frequency disturbances into dispatch problems mainly through stochastic processes, robust disturbance modeling, and chance-constrained formulations. Firstly, for the stochastic process approaches [24], the disturbances are modeled as continuous stochastic processes with temporal characteristics, such as stochastic differential equations [25] or autocorrelated random processes [26], to capture the uncertainty impacts on system frequency response. Secondly, for the robust optimization methods [27], the worst-case uncertainty sets are constructed to represent the frequency disturbances. In [28-29], the frequency security constraints are enforced to hold under all admissible realizations, which yields conservative dispatch decisions against adverse frequency impacts. Finally, for the chance-constrained formulations [30-31], the frequency security constraints are incorporated into dispatch models by enforcing probabilistic constraints on key frequency metrics such as RoCoF and frequency nadir. These formulations explicitly control the allowable risk of frequency violations induced by stochastic renewable fluctuations and contingencies [32]. However, existing studies do not account for spatially joint frequency disturbances arising from correlated stochastic renewable outputs across areas. Inaccurate disturbance characterization distorts frequency security constraints and leads to misestimated frequency risks in dispatch decisions.

To address the gaps above, this paper proposes a regional frequency-constrained dispatch method, which considers the spatial-joint stochastic disturbances and contingencies. The contributions of this paper are as follows:

1) A regional joint probabilistic frequency disturbance model is developed by modeling heterogeneous disturbances as the superposition of contingencies and renewable forecast-error-induced stochastic biases, with a Vine-Copula-based decomposition used to characterize their spatial dependence.

2) A Progressive Latent-Bottleneck Physics-Informed Neural Network (PLB-PINN) is developed to model the regional frequency dynamic process by embedding differential and integral representations of frequency response within a physics-informed encoder–decoder network model.

3) A day-ahead dispatch framework with multi-COIs frequency security constraints is formulated, co-optimizing unit commitment and regional primary reserves under spatial inertia heterogeneity and inter-regional support, with risk awareness enforced through conditional value-at-risk (CVaR)-based distribution-tail constraints.

## II. Framework

The proposed regional-frequency-constrained dispatch framework consists of four connected modules: stochastic disturbance modeling, multi-regional frequency response modeling, explicit representation of frequency security indicators, and day-ahead dispatch modeling, which is shown as Fig. 1.

Firstly, in the stochastic disturbance modeling module, regional equivalent active-power disturbances are constructed from renewable forecast errors and contingency-induced power deficits, and their spatial dependence is characterized using marginal distribution estimation and the Vine-Copula model. Secondly, in the frequency response modeling module, a multi-regional COI frequency dynamic model is established to describe regional frequency deviations, primary frequency response, and inter-regional tie-line power exchange on the second-level time scale, thereby deriving regional frequency security indicators. Thirdly, a PLB-PINN-based explicit representation is developed to characterize the high-order nonlinear relationship between dispatch decisions and regional frequency security indicators. The trained network is further reformulated into mixed-integer linear constraints and are embedded into the dispatch model. Finally, a regional frequency constrained dispatch model is constructed considering spatial-joint stochastic disturbances and contingencies, where CVaR constraints are introduced to limit the violation risk of regional frequency indicators under stochastic disturbances, so that sufficient regional frequency support can be reserved in advance. Through these modules, the proposed framework establishes a coordinated modeling connection between the day-ahead scheduling time scale and the second-level frequency response time scale, providing a unified solution for multi-regional frequency-security-constrained dispatch in high-renewable-penetration power systems.

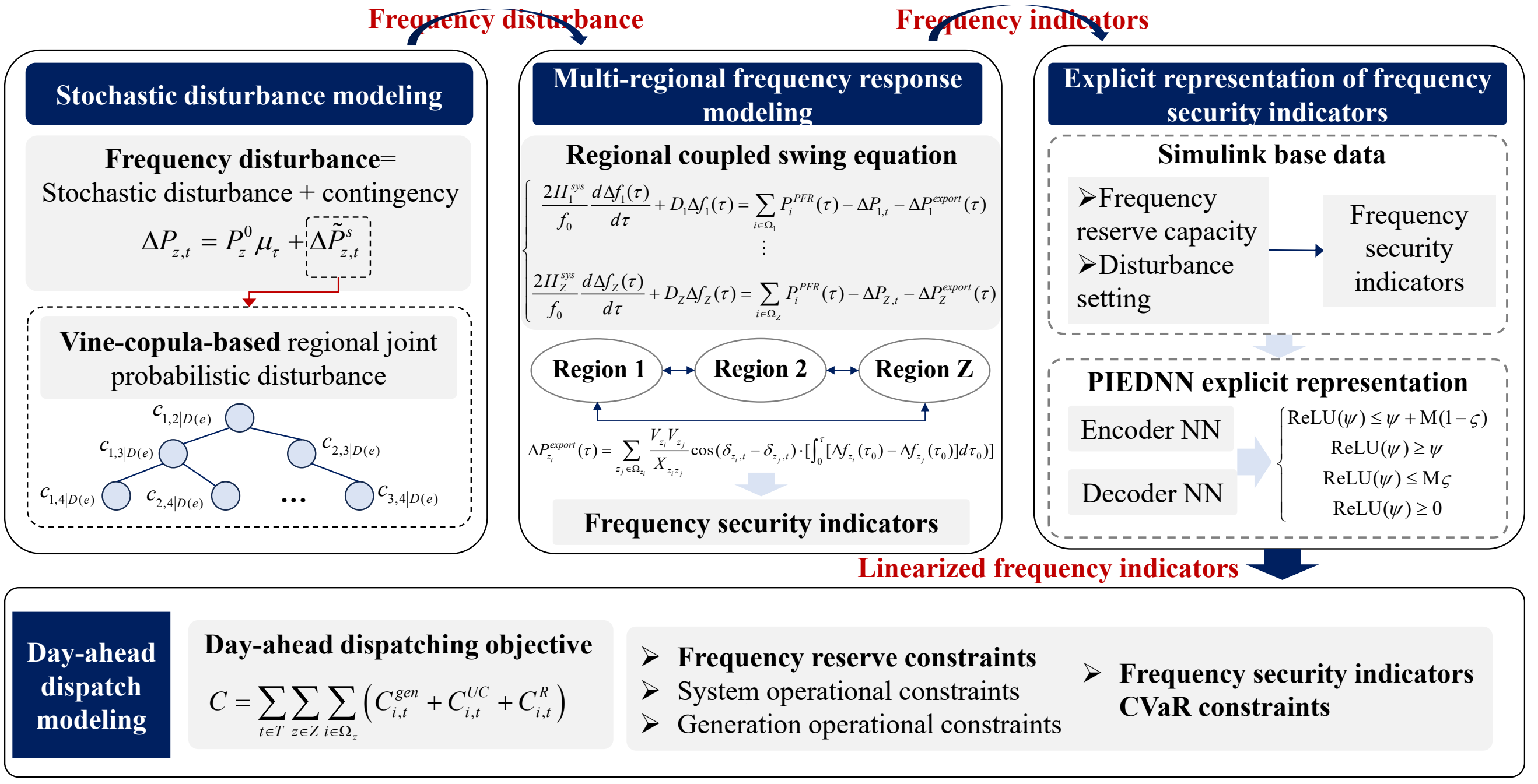

Fig. 1. Regional frequency constrained day-ahead dispatch framework considering spatial-joint stochastic disturbance and contingencies

## III. Regional Frequency Response Model Under Joint Distribution of Stochastic Disturbances

This section presents the formulation of regional frequency security indicator and the corresponding frequency response processes under stochastic disturbances. First, the equivalent stochastic disturbance is represented as the superposition of contingencies and the bias induced by renewable generation forecast errors. Next, the inter-regional dependence of these stochastic biases is modeled using a Vine Copula method. Finally, a coordinated regional frequency response architecture encompassing synchronous generators, WTs (Wind Turbines), and ESSs (Energy Storage Systems) is developed, based on which the regional frequency security indicator are derived. It is assumed that PV (Photovoltaic) is not included in this process because they have no rotating mechanical components and are mainly operated to provide active power generation to the grid. The regional frequency response model is shown in Fig. 2.

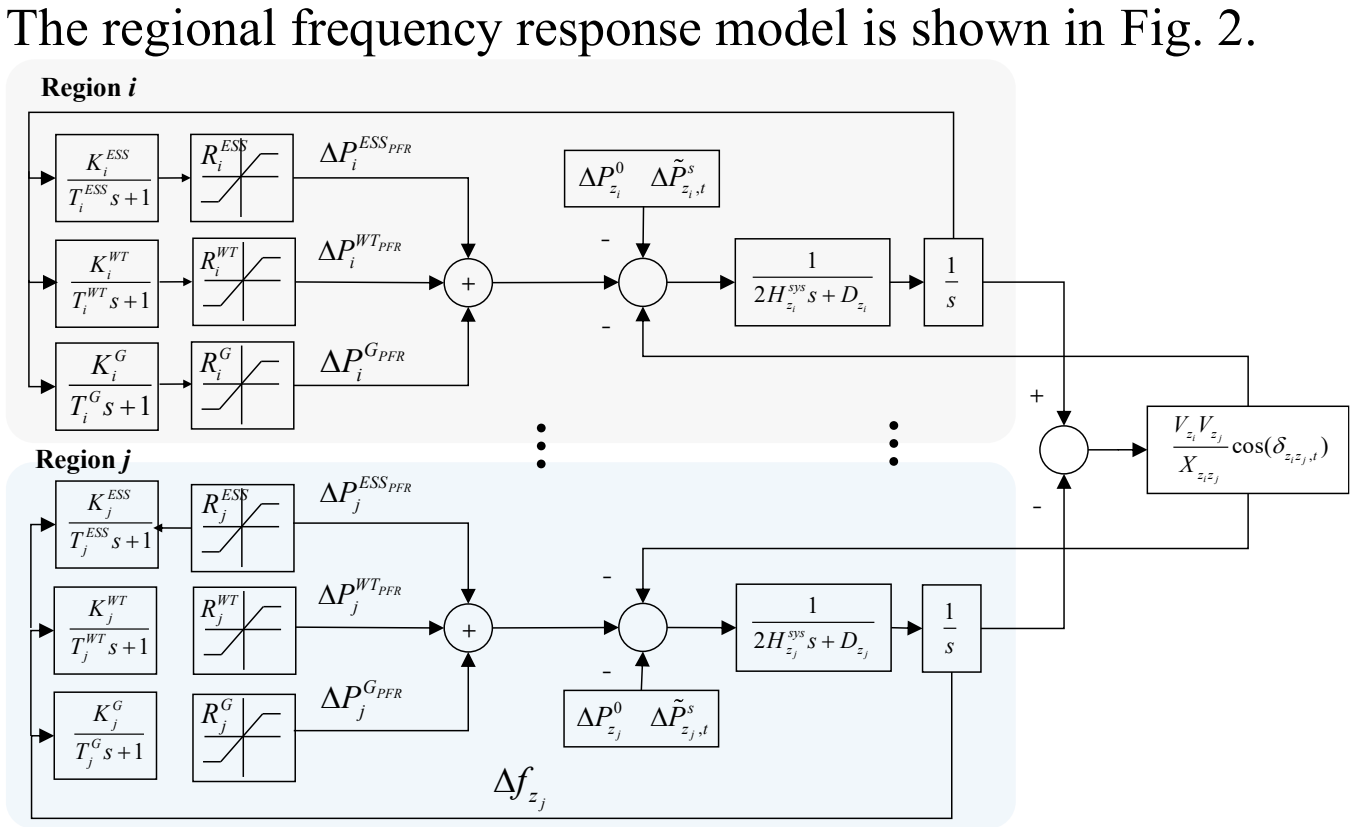

Fig. 2. Regional frequency supporting model

### A. Stochastic Disturbance Model

System frequency deviations originate from active power imbalances. Contingencies such as generator outages or sudden load losses introduce abrupt step changes in active power, leading to immediate frequency excursions. Also, forecast errors in renewable generation give rise to stochastic bias components, which manifest as persistent deviations between scheduled and actual power outputs. Accordingly, the equivalent active power disturbance can be expressed as:

$$\Delta P_{z,t} = P_z^0 \mu_\tau + \Delta \tilde{P}_{z,t}^s \tag{1}$$

For the frequency response model, the equivalent disturbance $\Delta P_{z,t}$ is introduced as the initial active power imbalance. This imbalance first causes regional frequency deviations, and then triggers primary frequency response and inter-regional power exchange.

The stochastic disturbance in region $z$ can be defined as the kernel density estimation:

$$f(\Delta \tilde{P}_{z,t}^s) \triangleq \frac{1}{Nh_z} \sum_{i=1}^{N} K\left( \frac{\Delta \tilde{P}_{z,t}^s - \Delta \tilde{P}_{z,t}^{s^{(i)}}}{h_z} \right) \tag{2}$$

where, $N$ is the number of stochastic disturbance samples. $\Delta \tilde{P}_{z,t}^{s(i)}$ is the $i$-th random sample of the stochastic disturbances.

Based on the PDF defined in Eq.(2), the marginal CDF is given by:

$$F(\Delta \tilde{P}_{z,t}^s) = \int_{-\infty}^{\Delta \tilde{P}_{z,t}^s} f(x)dx \tag{3}$$

$$\sum_{z}^{Z} F(\Delta \tilde{P}_{z,t}^s) = 1 \tag{4}$$

To characterize the spatial dependence of stochastic disturbances across multiple regions, a Vine Copula framework is adopted to construct the high-dimensional joint distribution. For a system with $Z$ regions, this framework consists of $Z$-1 hierarchical trees (denoted as $\{E_m\}$ where m=1,2,…,Z-1), forming a layered structure. Each tree $E_m$ contains multiple edges $e$, and each edge $e$ specifically connects two regions (indexed by $z_i$ and $z_j$). Notably, each edge $e$ is associated with a unique conditioning set D($e$) — this set D($e$) is exactly the subset of regions whose stochastic disturbances have been fully accounted for in the previous (i.e., prior to $E_m$) hierarchical trees. In turn, the dependence between the two regions $z_i$ and $z_j$ connected by edge $e$ is characterized under the condition of the disturbances from regions in D($e$).

The conditional dependence PDF between regions $z_i$ and $z_j$ under the conditioning set D($e$) is described by the associated pair-copula density $c_{zi,zj|D(e)}(\cdot)$. In this paper, the pair-copula density function is specified using a Gaussian copula, defined on the probability integral–transformed variables $u_z = F(\Delta \tilde{P}_{z,t}^s) \in (0,1)$.

$$\begin{aligned} c_{z_i,z_j|D(e)}(u_{z_i|D(e)}, u_{z_j|D(e)}) &= \Pr(u_{z_i} \le u_{z_i|D(e)}, u_{z_j} \le u_{z_j|D(e)}) \\ &= \Phi_{\rho_e}(\Phi^{-1}(u_{z_i|D(e)}), \Phi^{-1}(u_{z_j|D(e)})) \end{aligned} \tag{5}$$

Accordingly, the joint copula density of multi-regional disturbances is factorized into a product of conditional Gaussian pair-copula densities according to the chain rule of conditional densities.

$$c(u_1,...,u_Z) = \prod_{z=1}^{Z-1} \prod_{e \in E_m} c_{z_i,z_j|D(e)}(u_{z_i(e)|D(e)}, u_{z_j(e)|D(e)}) \tag{6}$$

With the joint copula density constructed above, the joint PDF of the multi-regional disturbances is obtained by combining the copula density with the corresponding marginal probability density functions according to Sklar's theorem.

$$f(\Delta \tilde{P}_1^s(t),\dots,\Delta \tilde{P}_Z^s(t)) = \left[ \prod_{z=1}^{Z} f(\Delta \tilde{P}_z^s(t)) \right] \cdot c(u_1,...,u_Z) \tag{7}$$

### B. Regional System Frequency Response Model

Based on the stochastic disturbances described above, the mathematical model of the multi-regional frequency response process is formulated as shown in Fig. 1. The regional coupled swing equation is expressed as:

$$\begin{cases} \frac{2H_1^{sys}}{f_0}\frac{d\Delta f_1(\tau)}{d\tau} + D_1\Delta f_1(\tau) = \sum_{i\in\Omega_1} P_i^{PFR}(\tau) - \Delta P_{1,t} - \Delta P_1^{export}(\tau) \\ \vdots \\ \frac{2H_Z^{sys}}{f_0}\frac{d\Delta f_Z(\tau)}{d\tau} + D_Z\Delta f_Z(\tau) = \sum_{i\in\Omega_Z} P_i^{PFR}(\tau) - \Delta P_{Z,t} - \Delta P_Z^{export}(\tau) \end{cases} \tag{8}$$

1) *Regional transmission power*

Due to heterogeneous disturbances and frequency support capabilities across different regions, phase angle differences arise following a frequency disturbance, leading to inter-regional oscillations and power exchanges. The regional transmission power deviations after disturbances modified the steady-state regional power balance and acted as an external disturbance for other regions.

$$\begin{aligned} \Delta P_{z_i}^{export}(\tau) &= \tilde{P}_{z_i}^{export}(\tau) - P_{z_i,t}^{export} \\ &= \sum_{z_j\in\Omega_{z_i}}\left[\frac{V_{z_i}V_{z_j}}{X_{z_iz_j}}\sin(\tilde{\delta}_{z_i}(\tau)-\tilde{\delta}_{z_j}(\tau)) - \frac{V_{z_i}V_{z_j}}{X_{z_iz_j}}\sin(\delta_{z_i,t}-\delta_{z_j,t})\right] \end{aligned} \tag{9}$$

The voltage amplitude and voltage phase angle are both obtained through the power flow at time slot *t*. $\Omega_{z_i}$ denotes the set of regions electrically adjacent to region $z_i$.

Considering small deviations around the operating point $(\delta_{zi} - \delta_{zj})$, the sine term can be linearized. Defining the incremental phase-angle disturbance as $\Delta\delta(\tau) = \tilde{\delta}(\tau) - \delta_t$, the term $\sin(\tilde{\delta}_{zi}(\tau) - \tilde{\delta}_{zj}(\tau))$ is approximated by performing a first-order Taylor expansion around the steady-state operating point $(\delta_{zi} - \delta_{zj})$. This linearization is adopted to obtain a tractable small-signal representation of the inter-regional power transfer under frequency disturbances, which facilitates integration into the frequency response equations.

$$\begin{aligned} &\sin(\tilde{\delta}_{z_i}(\tau)-\tilde{\delta}_{z_j}(\tau)) \approx \sin(\delta_{z_i,t}-\delta_{z_j,t}) \\ &+\cos(\delta_{z_i,t}-\delta_{z_j,t})[(\tilde{\delta}_{z_i}(\tau)-\tilde{\delta}_{z_j}(\tau))\text{-}(\delta_{z_i,t}-\delta_{z_j,t})] \end{aligned} \tag{10}$$

Then Eq. (9) can be transformed into:

$$\Delta P_{z_i}^{export}(\tau) = \sum_{z_j\in\Omega_{z_i}}\frac{V_{z_i}V_{z_j}}{X_{z_iz_j}}\cos(\delta_{z_i,t}-\delta_{z_j,t})[\Delta\delta_{z_i}(\tau)-\Delta\delta_{z_j}(\tau))] \tag{11}$$

The expression of the regional transmission power as the frequency deviation form is given by:

$$\frac{d\Delta\delta_z(\tau)}{d\tau} = \frac{d\tilde{\delta}_z(\tau)}{d\tau} - \frac{d\delta_{z,t}}{d\tau} \tag{12}$$

$$\frac{d\tilde{\delta}_z(\tau)}{d\tau} = \frac{d\theta_z(\tau)}{d\tau} - \omega_s = \omega_z(\tau) - \omega_s = 2\pi\Delta f_z(\tau) \tag{13}$$

At the fast time scale of the frequency response, the slow-time-scale operating point is assumed to be quasi-stationary, and thus the phase angle $\delta_{z,t}$ associated with the slow time scale satisfies $d\delta_{z,t}/d\tau = 0$.

Then the inter-regional transmission power deviations $\Delta P_z^{export}(\tau)$ can be expressed as:

$$\Delta P_{z_i}^{export}(\tau) = \sum_{z_j\in\Omega_{z_i}}\frac{V_{z_i}V_{z_j}}{X_{z_iz_j}}\cos(\delta_{z_i,t}-\delta_{z_j,t})\cdot[\int_0^\tau[\Delta f_{z_i}(\tau_0)-\Delta f_{z_j}(\tau_0)]d\tau_0)] \tag{14}$$

2) *Generator frequency response process*

In the inertial response process, inertia control replicates the rotor dynamic behavior of synchronous generators, WTs, and ESSs to deliver fast frequency support. The associated time-domain response model is expressed as:

$$P_i^{H_G/WT/ESS}(\tau) = -\frac{2H_i^{G/WT/ESS}}{f_0}\frac{d\Delta f_z(\tau)}{d\tau} \tag{15}$$

It is noted that each region is modeled using an equivalent COI to aggregate local inertial effects, whereas separate COIs are employed across regions to capture spatially heterogeneous inertial characteristics. The COI formulation is given as:

$$H_z^{sys} = \frac{\left[\sum_{i\in G_z}U_{i,t}^G H_i^G S_i^G + \sum_{i\in G_z}H_i^{WT}S_i^{WT} + \sum_{i\in G_z}H_i^{ESS}S_i^{ESS}\right]}{\sum_{i\in G_z}(U_{i,t}^G S_i^G + S_i^{WT} + S_i^{ESS})} \tag{16}$$

In the primary frequency response stage, droop control is employed to represent the primary frequency regulation behavior of synchronous generators, WTs, and ESSs. The resulting response process is expressed as:

$$P_i^{PFR_G/WT/ESS}(\tau) + T_i^{G/WT/ESS}\frac{dP_i^{PFR_G/WT/ESS}(\tau)}{d\tau} = -K_i^{G/WT/ESS}\Delta f_z(\tau) \tag{17}$$

The frequency security indicators are subsequently formulated based on the frequency response dynamics. $\text{RoCoF}_z$ is defined as the time derivative of the regional frequency in the region *z* after the disturbance defined in Eq. (1) occurs:

$$RoCoF_z(\tau) \triangleq \frac{d\Delta f_z(\tau)}{d\tau} \tag{18}$$

The frequency nadir refers to the minimum frequency reached after a disturbance, which characterizes the maximum frequency deviation from the nominal value. In this paper, the nadir point is described from two aspects: the construction of the training dataset for the PLB-PINN and the design of the physics-based regularization term. During the construction of the training dataset, the nadir value is obtained by directly searching for the minimum frequency value along the frequency response curve. During PLB-PINN training, the physical property that the frequency change rate equals zero at the nadir point is used to construct a regularization term, so as to enhance the network's capability to represent the physical characteristics of the nadir point. The description of the nadir point can be expressed as follows:

$$\begin{cases} f_z^{nadir} = \min(f) \\ \left.\frac{d\Delta f_z(\tau)}{d\tau}\right|_{f=f_z^{nadir}} = 0 \end{cases} \tag{19}$$

## IV. PLB-PINN FOR FREQUENCY SECURITY CONSTRAINTS

The frequency nadir formulated in Eq. (19) is governed by nonlinear differential dynamics, while the inter-regional supporting power in Eq. (14) is coupled through integral relationships. These coupled nonlinear characteristics render the frequency security constraints difficult to embed directly into the dispatch optimization problem.

To address this challenge, a PLB-PINN method is developed, in which the frequency nadir is treated as a causal latent variable linking dispatch decisions to frequency security. An encoder–decoder architecture is constructed, where the encoder maps high-dimensional dispatch variables to a nadir-centered latent representation, and the decoder reconstructs the associated physical responses, including primary frequency support and inter-regional power exchange, to enforce consistency with frequency dynamics. This design enhances physical interpretability and facilitates efficient embedding of frequency security constraints into the dispatch optimization.

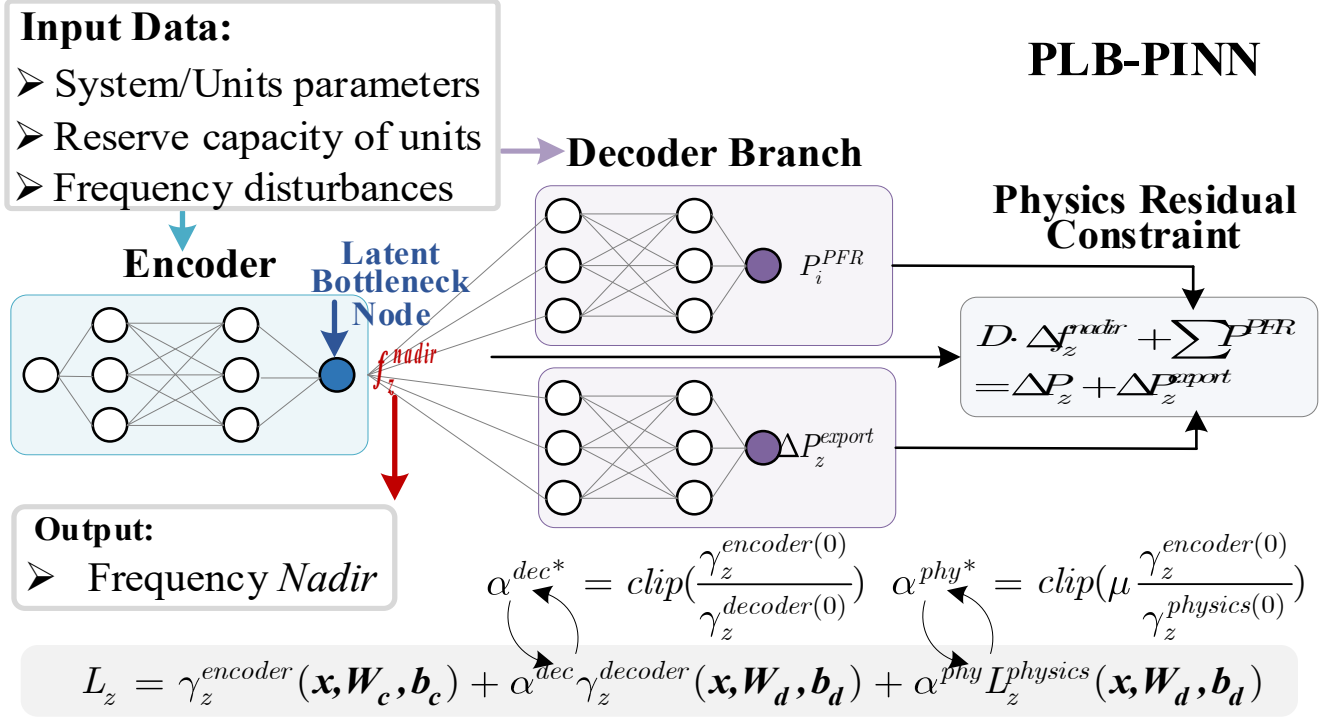

Fig. 3. The PLB-PINN architecture

### A. PLB-PINN Formulation

The encoder is implemented as a multilayer perceptron. Its input features consist of the primary frequency reserve capacities of synchronous generators, WTs, and ESSs.

The output of the encoder network includes the regional frequency nadir value $f_z^{nadir}$ and the network residual $\gamma_z^{encoder}$. The structure of the encoder network can be formulated as:

$$\hat{y}_z(f_z^{nadir}, \gamma_z^{encoder}) = f_z^{encoder}(\boldsymbol{x}, \boldsymbol{W}_c, \boldsymbol{b}_c) \tag{20}$$

$$\begin{gathered} f_z^{encader}(\boldsymbol{x}, \boldsymbol{W}_c, \boldsymbol{b}_c) = \boldsymbol{W}_{c3}\zeta_2 + \boldsymbol{b}_{c3} \\ \zeta_1 = \mathrm{ReLU}(\boldsymbol{W}_{c1}\boldsymbol{x} + \boldsymbol{b}_{c1}) \quad \zeta_2 = \mathrm{ReLU}(\boldsymbol{W}_{c2}\boldsymbol{\zeta}_1 + \boldsymbol{b}_{c2}) \end{gathered} \tag{21}$$

Within the PLB-PINN framework, the decoder takes the predicted frequency nadir as a latent bottleneck and maps it to the corresponding physical response variables, including the primary frequency response power $P_z^{PFR}$, the inter-regional transmission power $\Delta P_z^{export}$ at the nadir instant, and the network residual $\gamma_z^{decoder}$. The decoder mapping is expressed as:

$$\hat{y}_z(P_z^{PFR}, \Delta P_z^{export}, \gamma_z^{decoder}) = f_z^{decoder}(\boldsymbol{x}, \boldsymbol{W}_d, \boldsymbol{b}_d) \tag{22}$$

$$\begin{gathered} f_z^{decoder}(\boldsymbol{x}, \boldsymbol{W}_d, \boldsymbol{b}_d) = \boldsymbol{W}_{d3}\zeta_2 + \boldsymbol{b}_{d3} \\ \zeta_1 = \mathrm{ReLU}(\boldsymbol{W}_{d1}\boldsymbol{x} + \boldsymbol{b}_{d1}) \quad \zeta_2 = \mathrm{ReLU}(\boldsymbol{W}_{d2}\boldsymbol{\zeta}_1 + \boldsymbol{b}_{d2}) \end{gathered} \tag{23}$$

By substituting the dynamic trajectory predicted by the encoder into Eq. (8), the physics-informed regularization term is formulated as:

$$L_g^{physics}(\sigma_{PINN}) = \frac{1}{K}\sum_{k=1}^{K}\left|\begin{gathered}\frac{2H_z^{sys}}{f_0}\frac{d\Delta\hat{f}_z(\tau)}{d\tau} + D_z\Delta\hat{f}_z(\tau) - \sum_i P_{i\in G_z}^{PFR}(\tau) \\ + \Delta\hat{P}_z^{export}(\tau) + \Delta P_{z,t}\end{gathered}\right|^2 \tag{24}$$

Accordingly, the overall loss function of the PLB-PINN is composed of the encoder loss, the decoder loss, and a physics-informed regularization term:

$$\begin{aligned} L_z(\sigma_{PINN}) &= \alpha^{enc}\gamma_z^{encoder}(\boldsymbol{x}, \boldsymbol{W}_c, \boldsymbol{b}_c) \\ &+ \alpha^{dec}\gamma_z^{decoder}(\boldsymbol{x}, \boldsymbol{W}_d, \boldsymbol{b}_d) + \alpha^{phy}L_z^{physics} \end{aligned} \tag{25}$$

To promote stable joint training of the decoder outputs and the embedded physical constraints, an adaptive weighting mechanism is introduced to automatically balance the data-driven reconstruction loss against the physics-informed regularization term.

Firstly, the initial loss weights for each component are defined as $\alpha^{enc(0)}$, $\alpha^{dec(0)}$, $\alpha^{phy(0)}$. Then the targeted weight for the physics network is:

$$\alpha^{phy*} = \gamma\frac{\alpha^{enc(0)} * \alpha^{dec(0)}}{\alpha^{phy(0)} + \varepsilon} \tag{26}$$

where $\gamma \ll 1$ is a scaling factor introduced to suppress the influence of the physics-based terms during the early stages of training, and $\varepsilon$ is a numerical stabilization parameter used to avoid division by zero.

To further enhance training robustness, the physics-informed regularization is progressively activated through a ramp-up scheduling strategy. Specifically, a monotonic activation factor is introduced as:

$$\lambda_e = \min(1, \frac{ep}{E^{ep}}) \tag{27}$$

Then the weight for the physics-based regularization term is:

$$\alpha^{phy} = \alpha^{phy*}\lambda_e \tag{28}$$

### B. Frequency Security Constraints Linearization

Since the PLB-PINN is a highly nonlinear module, it cannot be directly embedded into the system dispatching optimization problem. To address this, all activation functions in the network are replaced with ReLU functions, whose piecewise linear characteristics allow for exact linearization using the Big-M method [20]. This enables the network to be incorporated into a tractable mixed-integer linear programming optimization model. The ReLU activation function is defined as:

$$\mathrm{ReLU}(\psi) = \mathrm{Max}(0, \psi) \tag{29}$$

By introducing a binary variable $\varsigma \in \{0,1\}$ and a sufficiently large constant M, the ReLU function can be equivalently reformulated as the following set of mixed-integer linear constraints:

$$\begin{cases} \mathrm{ReLU}(\psi) \le \psi + \mathrm{M}(1-\varsigma) \\ \mathrm{ReLU}(\psi) \ge \psi \\ \mathrm{ReLU}(\psi) \le \mathrm{M}\varsigma \\ \mathrm{ReLU}(\psi) \ge 0 \end{cases} \tag{30}$$

Since only the encoder outputs (i.e., frequency security indicator) are embedded into the dispatch optimization model,

the linearization is performed exclusively on the encoder. The decoder serves as an auxiliary module to enhance frequency trajectory reconstruction and is not involved in the optimization, as its outputs do not affect dispatch decisions.

## V. Regional Day-ahead Dispatching Model Under Stochastic Frequency Security Constraints

### A. Model Formulation

The objective of the day-ahead dispatching scheduling optimization problem is to minimize the day-ahead total cost.

$$C=\sum_{t\in T}\sum_{z\in Z}\sum_{i\in\Omega_z}\left(C_{i,t}^{gen}+C_{i,t}^{UC}+C_{i,t}^{R}\right) \tag{31}$$

$$\begin{cases} C_{i,t}^{gen}=U_{i,t}^{G}\cdot(a_i\cdot(P_{i,t}^{G})^2+b_i\cdot P_{i,t}^{G}+c_i) \\ C_{i,t}^{UC}=\lambda_i^{start}\cdot x_{i,t}^{G_on}+\lambda_i^{off}\cdot x_{i,t}^{G_off} \\ C_{i,t}^{R}=\lambda_i^{R_th}\cdot R_{i,t}^{G}+\lambda_i^{R_WT}\cdot R_{i,t}^{WT}+\lambda_i^{R_ESS}\cdot R_{i,t}^{ESS} \end{cases} \tag{32}$$

The dispatching model should be subject to the following constraints:

1) *CVaR-based Stochastic Frequency Security Constraints:*

To embed stochastic frequency security constraints into the day-ahead dispatch model, the CVaR is employed as a convex risk measure, providing a probabilistic and optimization-tractable representation of frequency security under uncertainty.

To quantify frequency violation risks, a loss function is defined to measure the extent of violations relative to the prescribed security threshold:

$$\begin{cases} L_z^{RoCoF(s)}=\max\left(\max(RoCoF_z^{(s)}(\tau))-\overline{RoCoF},0\right) \\ L_z^{nadir(s)}=\max\left(f_z^{nadir(s)}-\overline{f^{nadir}},0\right) \end{cases} \tag{33}$$

For a specified confidence level $\beta$, the value-at-risk, denoted by $\eta_z$, defines a threshold such that the probability of the loss exceeding $\eta_z$ is no greater than $1-\beta$. To obtain a convex representation of CVaR, the standard Rockafellar–Uryasev formulation is adopted [34]. Introducing slack variables $\xi_z^{RoCoF(s)}$ and $\xi_z^{nadir(s)}$, the CVaR is formulated as:

$$\begin{cases} CVaR_z^{RoCoF}(\beta)=\eta_z^{RoCoF}+\dfrac{1}{1-\beta}\sum_{s\in\Omega_s}\dfrac{1}{|\Omega_s|}\xi_z^{RoCoF(s)} \\ CVaR_z^{nadir}(\beta)=\eta_z^{nadir}+\dfrac{1}{1-\beta}\sum_{s\in\Omega_s}\dfrac{1}{|\Omega_s|}\xi_z^{nadir(s)} \end{cases} \tag{34}$$

$$\begin{cases} \xi_z^{RoCoF(s)}\ge\max(L_z^{RoCoF(s)}-\eta_z^{RoCoF},0) \\ \xi_z^{nadir(s)}\ge\max(L_z^{nadir(s)}-\eta_z^{nadir},0) \end{cases} \tag{35}$$

It is worth noting that Eq. (33) and Eq. (35) can be linearized and incorporated into the dispatch problem using the big-M method [20].

Then the stochastic frequency security constraints are expressed as:

$$\begin{cases} CVaR_z^{RoCoF}(\beta)\le\varepsilon^{RoCoF} \\ CVaR_z^{nadir}(\beta)\le\varepsilon^{nadir} \end{cases} \tag{36}$$

2) *System operational constraints:*

Regional power balance should be maintained:

$$\begin{aligned} &\sum_{i\in G_z}P_{i,t}^{G}+\sum_{i\in G_z}P_{i,t}^{re}+\sum_{i\in G_z}(P_{i,t}^{dis}-P_{i,t}^{ch}) \\ &=\sum_{i\in G_z}P_{i,t}^{load}+P_{z,t}^{export} \end{aligned} \tag{37}$$

$$\sum_{z_i\in\Omega_Z}\sum_{z_j\in\Omega_{z_i}}P_{z_iz_j,t}^{export}=0 \tag{38}$$

The power system is modeled using a DC optimal power flow formulation, in which each region is represented by an equivalent node. The resulting model is expressed as:

$$\underline{P}_{z_iz_j}\le P_{z_iz_j,t}^{export}\le\overline{P}_{z_iz_j} \tag{39}$$

3) *Generators Operation Constraints:*

Each synchronous generator must operate within its allowable generation range:

$$U_{i,t}^{G}\underline{P_i^{G}}\le P_{i,t}^{G}+R_{i,t}^{G}\le U_{i,t}^{G}\overline{P_i^{G}} \tag{40}$$

Generators must also comply with ramp-up and ramp-down limitations:

$$\begin{cases} P_{i,t}^{G}-P_{i,t-1}^{G}+R_{i,t}^{G}-R_{i,t-1}^{G}\le R_i^{UP} \\ P_{i,t-1}^{G}-P_{i,t}^{G}+R_{i,t-1}^{G}-R_{i,t}^{G}\le R_i^{DN} \end{cases} \tag{41}$$

The start-up and shut-down constraints are shown as follows:

$$U_{i,t}^{G}-U_{i,t-1}^{G}=x_{i,t}^{G_on}-x_{i,t}^{G_off} \tag{42}$$

$$\begin{cases} \sum_t^{\min(T,t+t_i^{on}-1)}U_{i,t}^{G}\ge t_i^{on}\cdot x_{i,t}^{G_on} \\ \sum_t^{\min(T,t+t_i^{off}-1)}(1-U_{i,t}^{G})\ge t_i^{off}\cdot x_{i,t}^{G_off} \end{cases} \tag{43}$$

The charging and discharging power of ESSs, the capacity of ESSs must satisfy the maximum output/capacity limitations:

$$P_{i,t}^{dis}/\eta_i^{dis}-P_{i,t}^{ch}\eta_i^{ch}+R_i^{ESS}\le\overline{P_i^{ESS}} \tag{44}$$

$$E_{i,t}^{ESS}=E_{i,t-1}^{ESS}-P_{i,t}^{dis}/\eta_i^{dis}+P_{i,t}^{ch}\eta_i^{ch} \tag{45}$$

$$E_{i,T}^{ESS}=E_{i,0}^{ESS} \tag{46}$$

$$\underline{E_{i,t}^{ESS}}\le E_{i,t}^{ESS}-R_i^{ESS}\Delta t\le\overline{E_{i,t}^{ESS}} \tag{47}$$

The output of renewable energy must satisfy the minimum and maximum output limits:

$$\begin{aligned} 0\le P_{i,t}^{wt}+R_{i,t}^{WT}\le\overline{P_i^{wt}} \\ 0\le P_{i,t}^{pv}\le\overline{P_i^{pv}} \end{aligned} \tag{48}$$

## VI. Case Study

The day-ahead dispatching method, considering the regional frequency security indicator and the spatial-joint stochastic disturbance, is tested. The proposed dispatching optimization model is addressed using Gurobi 11.0.0, while the PLB-PINN is programmed utilizing Python 3.10 on a computer equipped with a 2.4 GHz CPU (i7-13620H), an Intel (R) UHD graphics GPU, and 16GB RAM.

### A. Basic Data

The proposed method is validated on the modified IEEE 118-bus system [35], where three regions are divided according to the geographic location and the blocking section, which is shown in Fig. 4. Each region is assumed under a unified COI, with inter-regional frequency support. Among the three regions, Region 1 is dominated by thermal generation with relatively high synchronous inertia and primary frequency response capability. Region 2 represents a renewable-energy base with a higher wind penetration level. Region 3 is a load center with great demand, which relies on the other regions' power support.

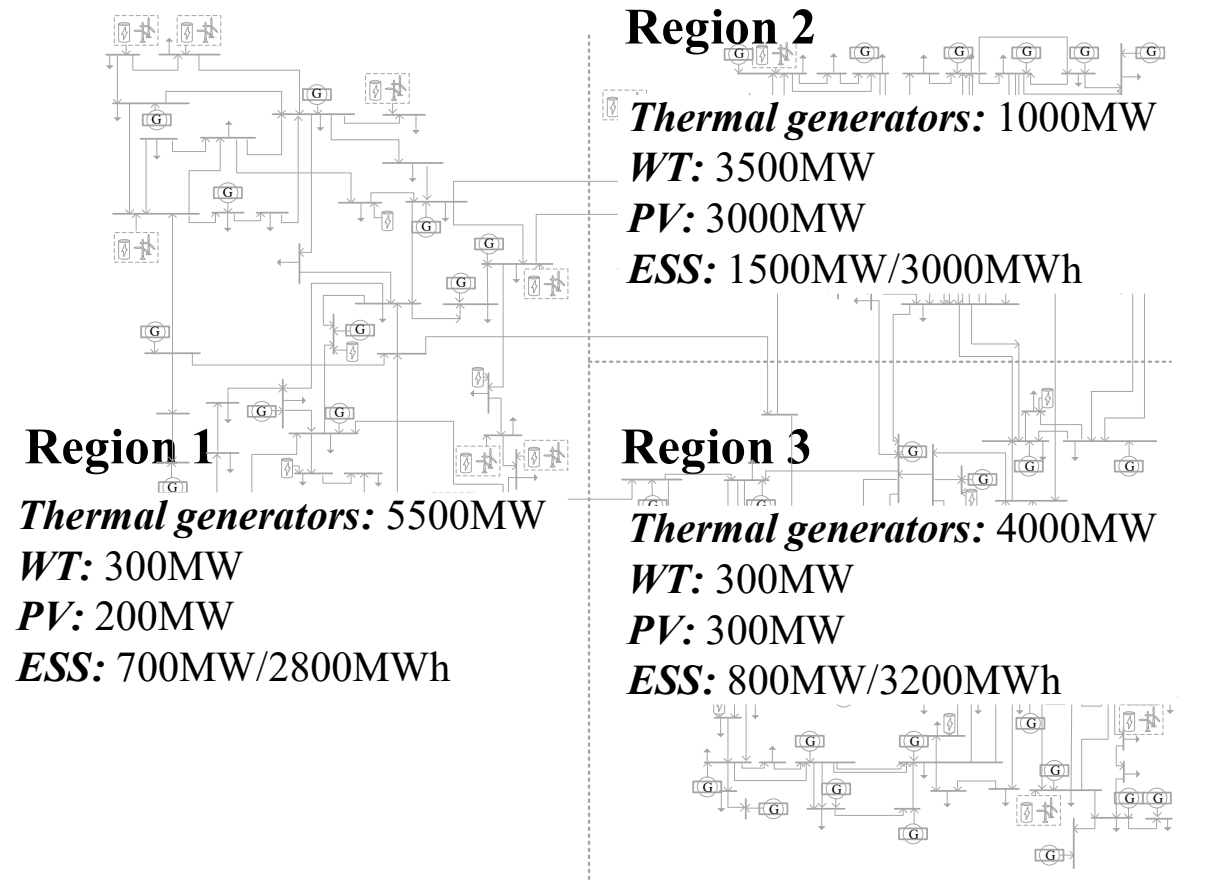

Fig. 4. The topology and units' capacity of the modified IEEE-118 bus system.

System disturbances are modeled as the superposition of a system failure contingency and stochastic renewable energy output forecast errors. The system failure is modeled as a 10% power deficit of the regional maximum load and is assumed to occur in Region 3, since its largest load level results in the greatest absolute power deficit and the most significant impact on system frequency. Renewable uncertainty is modeled by day-ahead forecast errors with a 15-min time resolution. For each region, the forecast error is assumed to follow a Gaussian distribution with standard deviations set to 10% in Region 1, 15% in Region 2, and 8% in Region 3 of the rated renewable output.

### B. *Result of the day-ahead dispatching optimization*

#### 1) *Dispatching schedule*

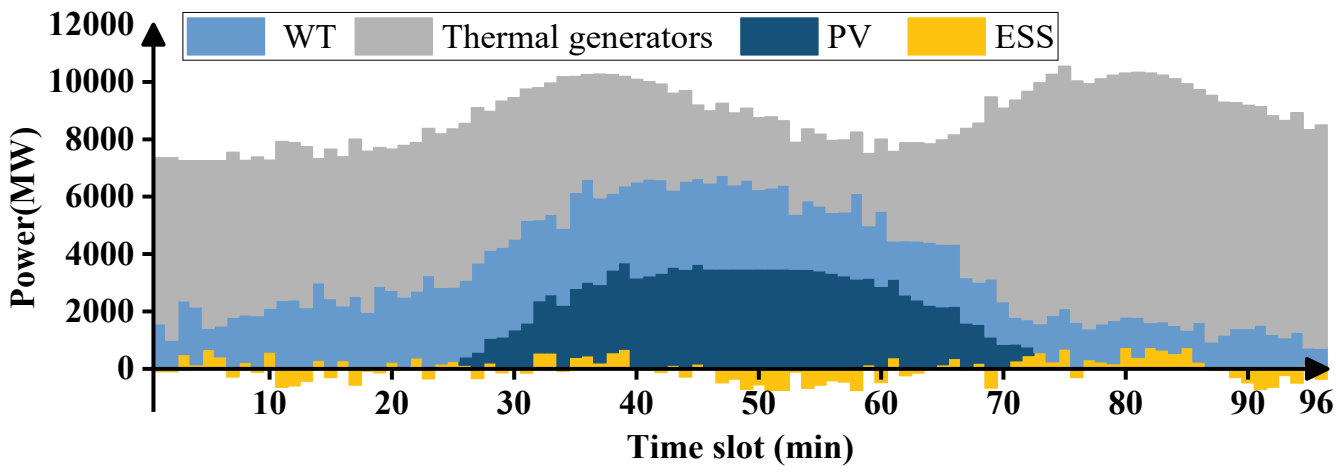

Fig. 5. The day-ahead dispatching schedule

Fig. 5 illustrates the day-ahead dispatch results of the proposed model. During the time slot 40 to 52, the renewable energy generation is much higher, accounting for 80.33% to 88.80% of the installed renewable capacity. Two load peak periods are observed during time slots 36-44 and 75-85. In the first peak period, the high demand is primarily supplied by abundant WT generation, whereas in the second peak period, reduced WT and PV outputs require thermal generators to operate near their upper limits to meet the evening peak demand. In both peak intervals, ESSs are dispatched in discharging mode to provide additional support and smooth the net load profile. During the midday period with high PV generation, ESSs are charged to absorb surplus renewable energy and enable inter-temporal energy shifting.

#### 2) *Frequency support schedule*

Fig. 6 illustrates the total allocated primary frequency reserves in each time slot and the thermal generators, WT, and ESSs reserve capacities in each region. Taking time slots 30-45 and 75-85 as examples, both intervals have similar load levels, while the renewable output is higher during time slots 30-45. Accordingly, the average scheduled primary frequency reserve reaches 2063.98 MW during time slots 30-45, which is higher than the 1424.78 MW during time slots 75-85. In contrast, during time slots 35-40 and 55-60, the renewable outputs are close, but the load demand is higher during time slots 35-40. As a result, the average scheduled reserve reaches 2019.90 MW during time slots 35-40, compared with 1800.62 MW during time slots 55-60. These results indicate that reserve allocation is not determined by load demand alone, but by the combined effects of load level, renewable forecast-error uncertainty, and the resulting equivalent frequency disturbance.

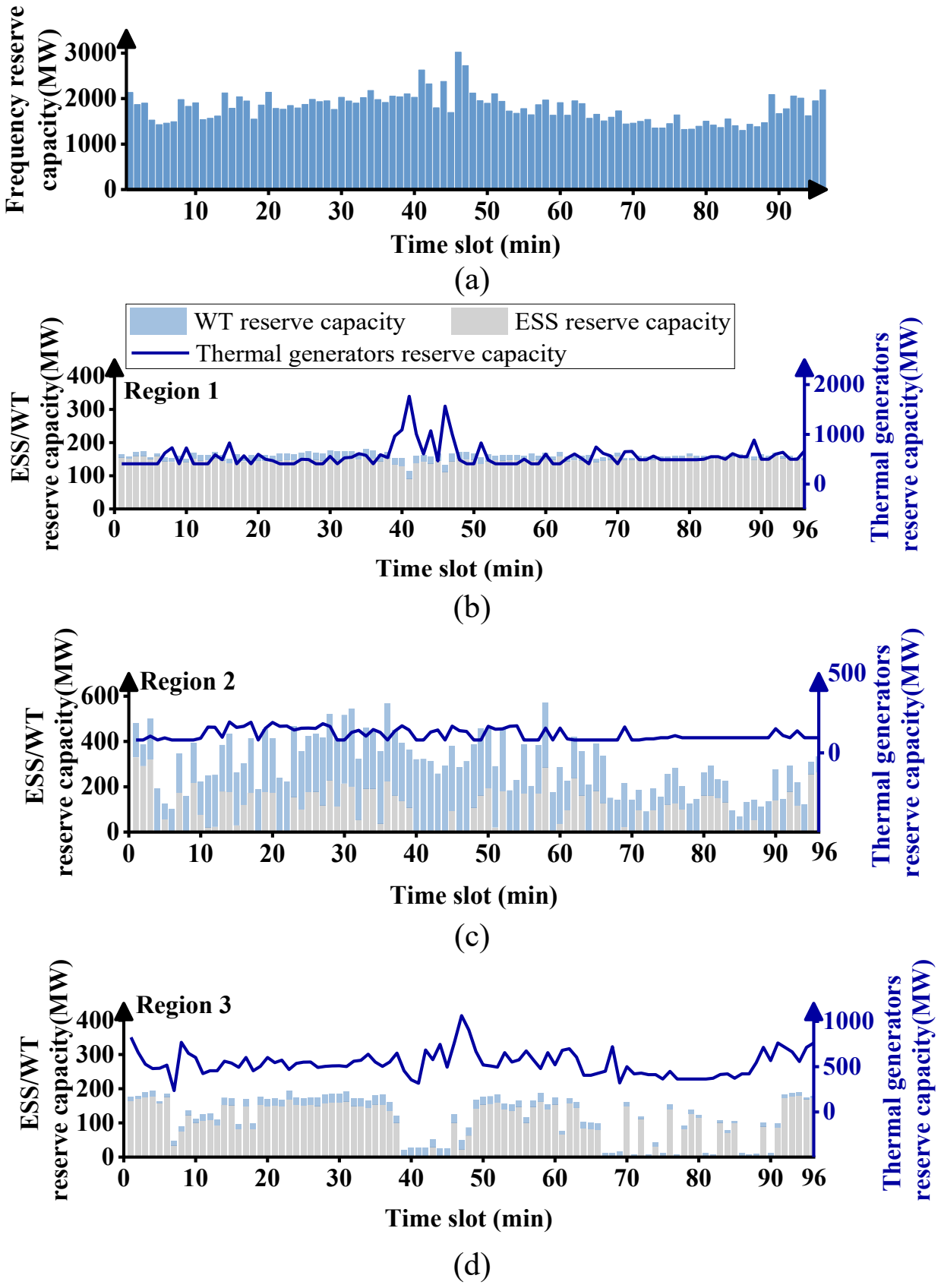

Fig. 6. Total frequency reserve capacity(a) and frequency reserve capacity of thermal generators, WT, and ESSs in each region (b), (c), (d)

Fig. 7 shows the corresponding nadir and RoCoF margins. **From the spatial perspective**, Region 1 allocates a large amount of thermal primary frequency reserve of 563.79 MW, while WT and ESS reserves remain relatively low at 14.75 MW and 147.30 MW, indicating that frequency support is mainly provided by thermal units. In Region 2, WT-based reserves dominate the reserve composition, averaging 200.71 MW, supplemented by ESS reserves of 99.27 MW to mitigate renewable variability. By contrast, Region 3 relies predominantly on ESS-based reserves, with an average of 103.06 MW and notable temporal variations, despite sufficient thermal capacity being available. Overall, the heterogeneous reserve allocation across regions reflects differences in generation portfolios, response characteristics, and local frequency constraints.

**From the temporal perspective,** the frequency security indicators deteriorate most significantly during time slots 36-52 The frequency nadir margin drops to -0.32 Hz in the average of Region 3, and the RoCoF increases to -0.33 Hz/s, representing the worst performance over the entire horizon. This time interval coincides with intensive energy production from thermal units, WTs, and ESSs, leading to the lowest available primary frequency reserve, which decreases to about 240.09 MW. Meanwhile, the high WT output during this period increases the stochastic error, resulting in larger equivalent disturbances. As a result, the reduced reserve availability combined with increased renewable uncertainty disturbs during time slots 36-52, leading to the most critical frequency security conditions.

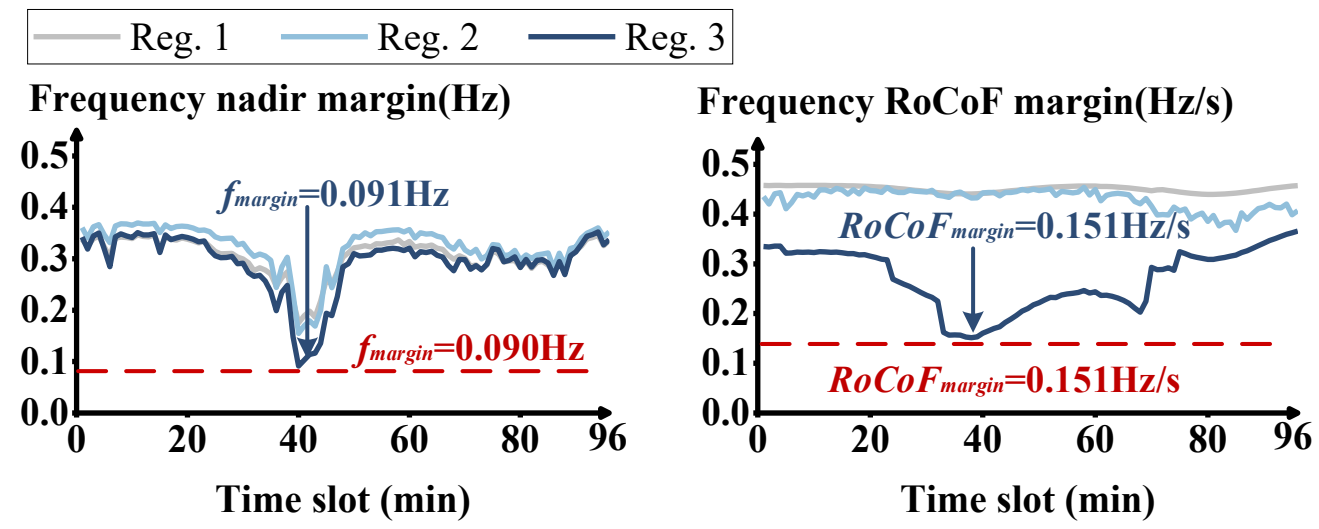

Fig. 7. The frequency nadir and RoCoF indicator margin in each region

3) *Frequency response process*

To clarify how the reserved frequency support is actually deployed after the disturbance, Fig. 8 illustrates the generators and transmission-line response process at time slot 40 as an example.

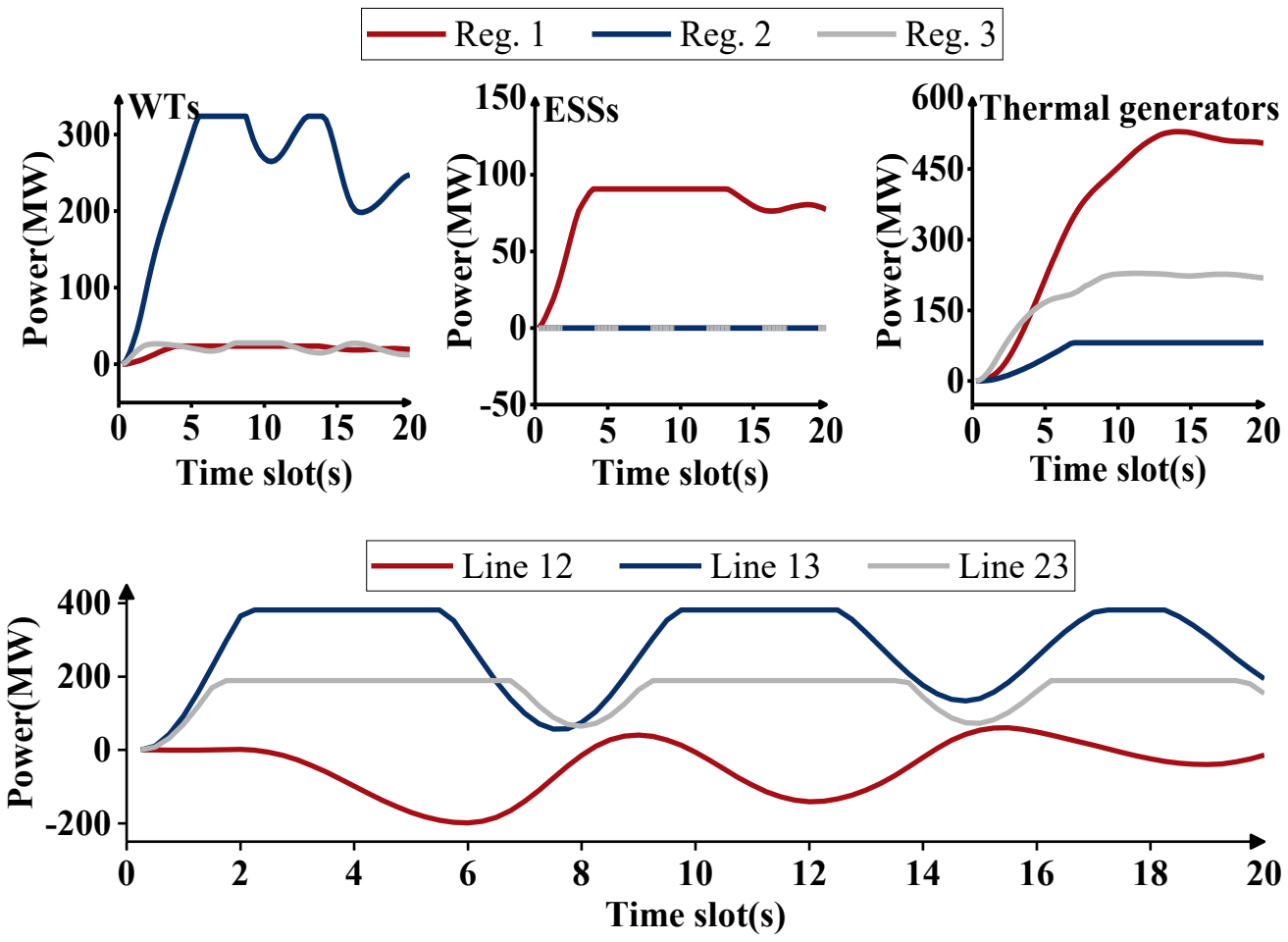

Fig. 8. The generators and transmission-line response process of the paper proposed case in T=40

Under operating condition setting above, the three regions reserve different levels of primary frequency response capacity (Region 1: 1880.71MW, Region 2: 404.94MW, and Region 3: 345.19MW) according to their local frequency support capability and inter-regional transmission constraints. The WT frequency supporting is mainly provided by Region 2, whose peak WT frequency response power reaches 323.64 MW from 5.5-8.75s after disturbance, which is much higher than those of Regions 1 and 3. By contrast, Region 1 provides the largest thermal-generator and ESS power responses, with peak values of 526.52MW and 90.65MW, respectively. Meanwhile, the inter-regional tie-line power exchange varies dynamically after the disturbance. Line 13 and Line 23 provide considerable inter-regional support, whereas Line 12 exhibits bidirectional power fluctuation. This indicator that the frequency response of Region 3 is jointly supported by local frequency response resources and external resources delivered from neighboring regions through inter-regional tie-lines.

### C. *Analysis of regional COI representation*

To evaluate the effectiveness of the proposed regional frequency supporting framework, a conventional unified COI–based dispatching model is applied:

**Regional COIs:** The proposed day-ahead dispatching method with regional-specific frequency security constraints.

**Unified COI:** The day-ahead dispatching method based on a unified COI frequency model.

1) *Dispatching schedule*

Fig. 9 compares the day-ahead dispatching schedules under time slots 36–52, which correspond to the most critical frequency security period. Table I compares the day-ahead scheduling costs obtained by the regional COI-based method and the unified COI-based method across regions.

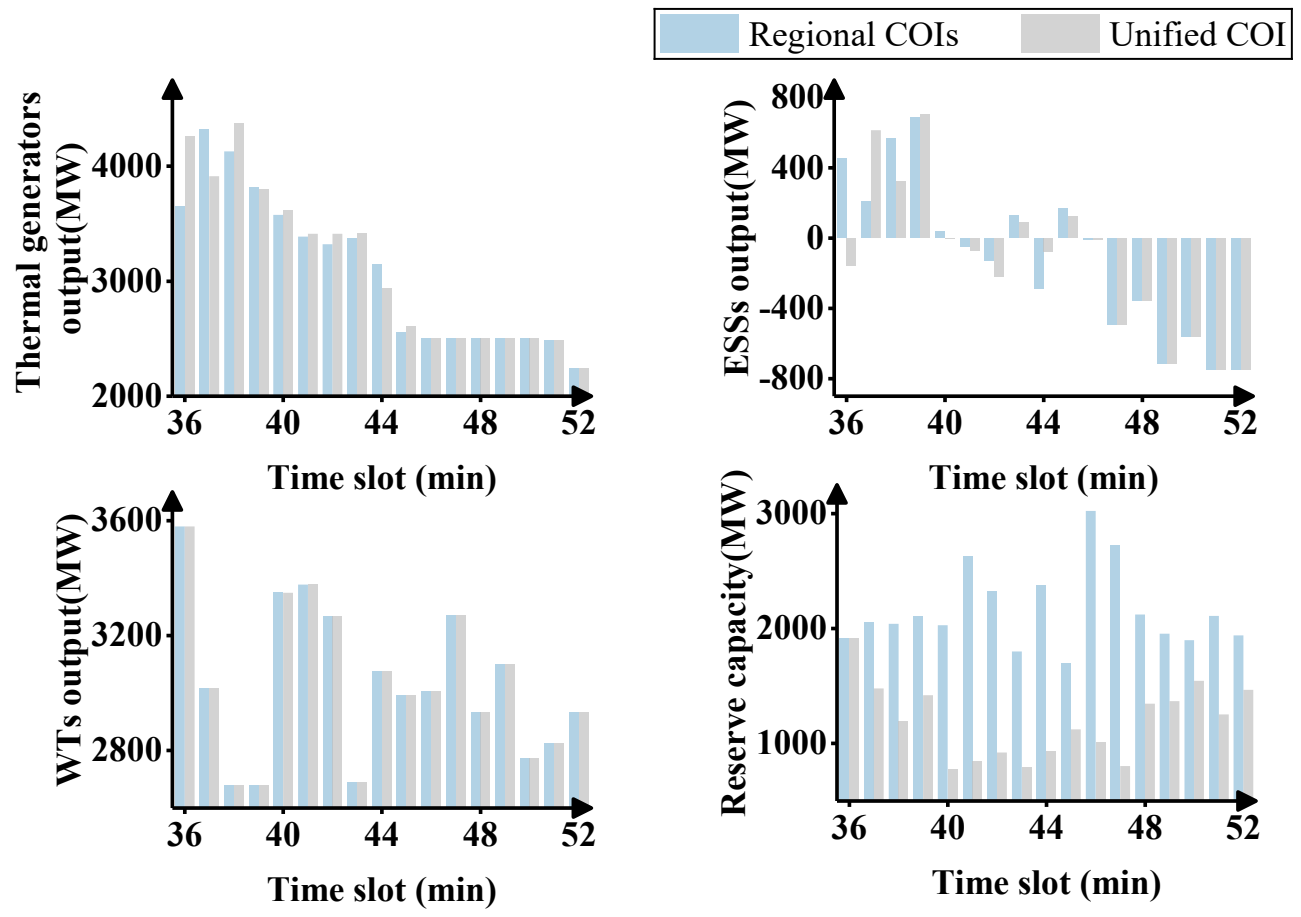

Fig. 9. Day-ahead dispatching schedule of regional COIs and unified COI

TABLE I
THE COMPARISON OF THE COSTS BETWEEN THE REGIONAL COIS AND THE UNIFIED COI ACROSS REGIONS

| Case / Costs( $ ) | Regional COIs | | | Unified COI | | |
|---|---|---|---|---|---|---|
| | Reg. 1 | Reg. 2 | Reg. 3 | Reg. 1 | Reg. 2 | Reg. 3 |
| **Generation costs** ($\times10^6$) | 1.98 | 0.33 | 1.25 | 1.72 | 0.31 | 1.38 |
| **Reserve costs** ($\times10^6$) | 0.70 | 0.40 | 0.63 | 0.69 | 0.53 | 0.59 |
| **UC costs** ($\times10^6$) | 0.05 | 0.01 | 0.08 | 0.05 | 0.01 | 0.08 |
| **Total costs** ($\times10^6$) | 2.73 | 0.74 | 1.96 | 2.46 | 0.85 | 2.05 |

The total operating cost of the proposed method is $5.43 million, compared with $5.36 million under the COI-based method, representing an increase of 1.31%. Specifically, the proposed method results in a 4.40% increase in generation cost and a 4.41% reduction in reserve cost. The higher generation cost mainly stems from different dispatch patterns. By enforcing regional frequency security constraints, thermal units in frequency-critical regions are dispatched away from their most cost-efficient operating points to maintain adequate frequency support. In particular, additional reserves are retained in Region 3 to address its tighter local frequency constraints. Overall, the proposed method improves regional frequency security with only a modest increase in total cost through a more targeted reserve allocation.

2) *Regional support and frequency performance improved*

Fig. 10 and Fig. 11 compare the post-disturbance frequency response process at time slot T=40 under the proposed regional COIs method and the unified COI-based method, respectively.

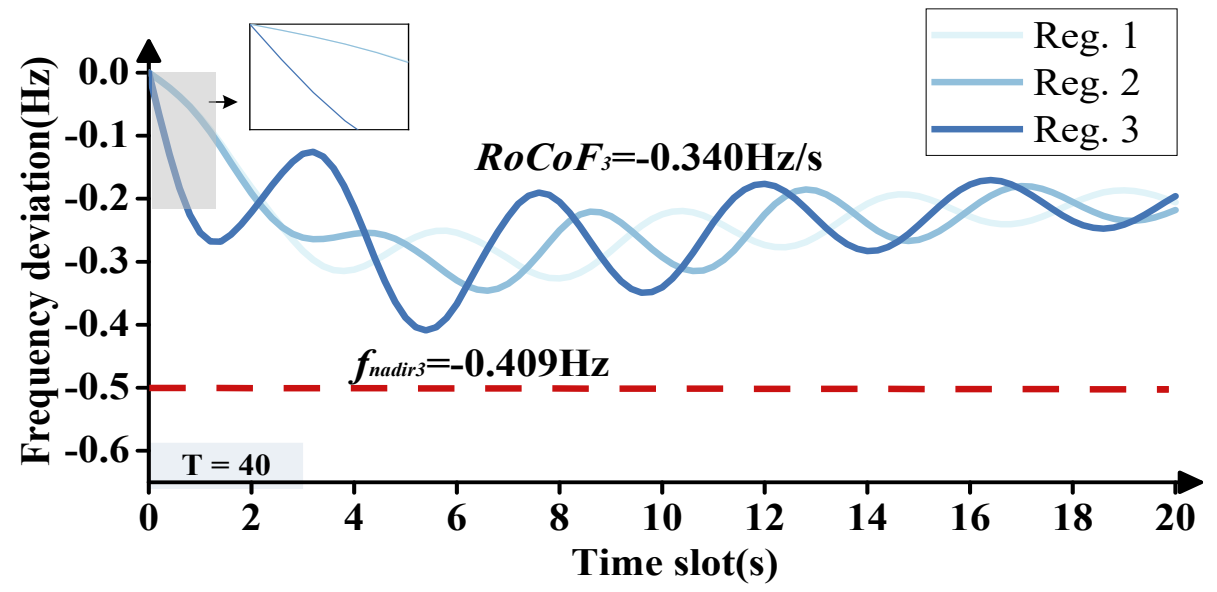

Fig. 10. The frequency dynamic process of the paper proposed case in T=40

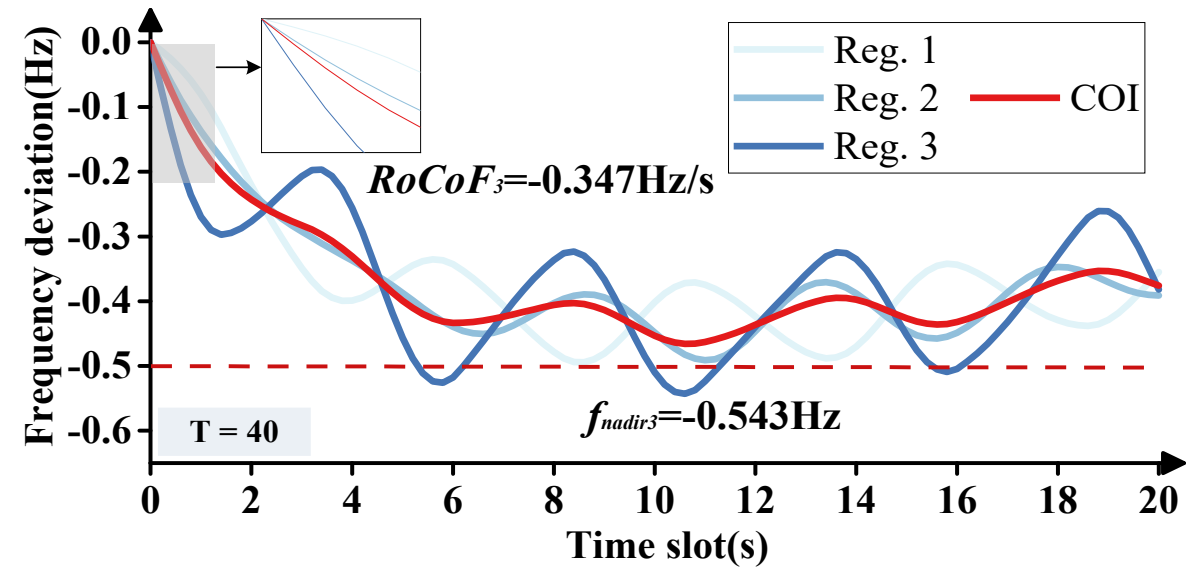

Fig. 11. The frequency dynamic process of the COI case in T=40

As shown in Fig. 10, under the proposed Regional COIs method, both the RoCoF and nadir indicator remain within the prescribed limits in all three regions. Among them, Region 3 exhibits the most critical response, with the worst nadir and RoCoF values of -0.409Hz and -0.34Hz/s, respectively, which is consistent with its weaker local frequency strength and higher sensitivity to disturbances. Fig. 11 indicates that under the Unified COI-based method, although the aggregated system-level frequency indicator do not violate the limits, the Region 3 nadir exceeds the security threshold, reaching -0.543Hz. This discrepancy arises because the unified COI representation averages spatial frequency dynamics and may mask localized frequency stress, leading to reserve and support decisions that appear sufficient at the system level but are insufficient for the most vulnerable region.

## D. *Analysis of the stochastic disturbance*

To evaluate the effectiveness of the proposed spatial-joint distribution of stochastic disturbances, a spatial dispersed distribution–based dispatching model is applied:

**Spatial-joint disturbances:** The proposed day-ahead dispatching method considers the spatial-joint distribution of stochastic disturbances.

**Spatial-dispersed disturbances:** The day-ahead dispatching method considers the dispersed distribution of the stochastic disturbance.

1) *Frequency performance and dispatching schedule*

Fig. 12 compares the frequency security performance, and Fig. 13 shows the corresponding day-ahead dispatching schedules under the spatial-joint distribution and the dispersed distribution of the stochastic disturbance case.

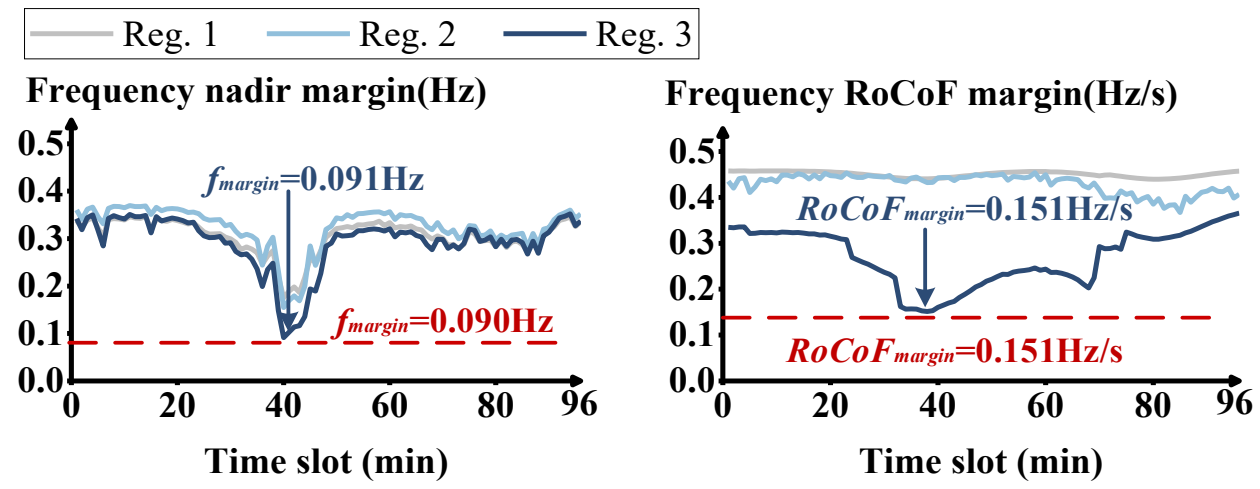

Fig. 12. The frequency nadir and RoCoF indicator margin in each region under the dispersed distribution of the stochastic disturbance

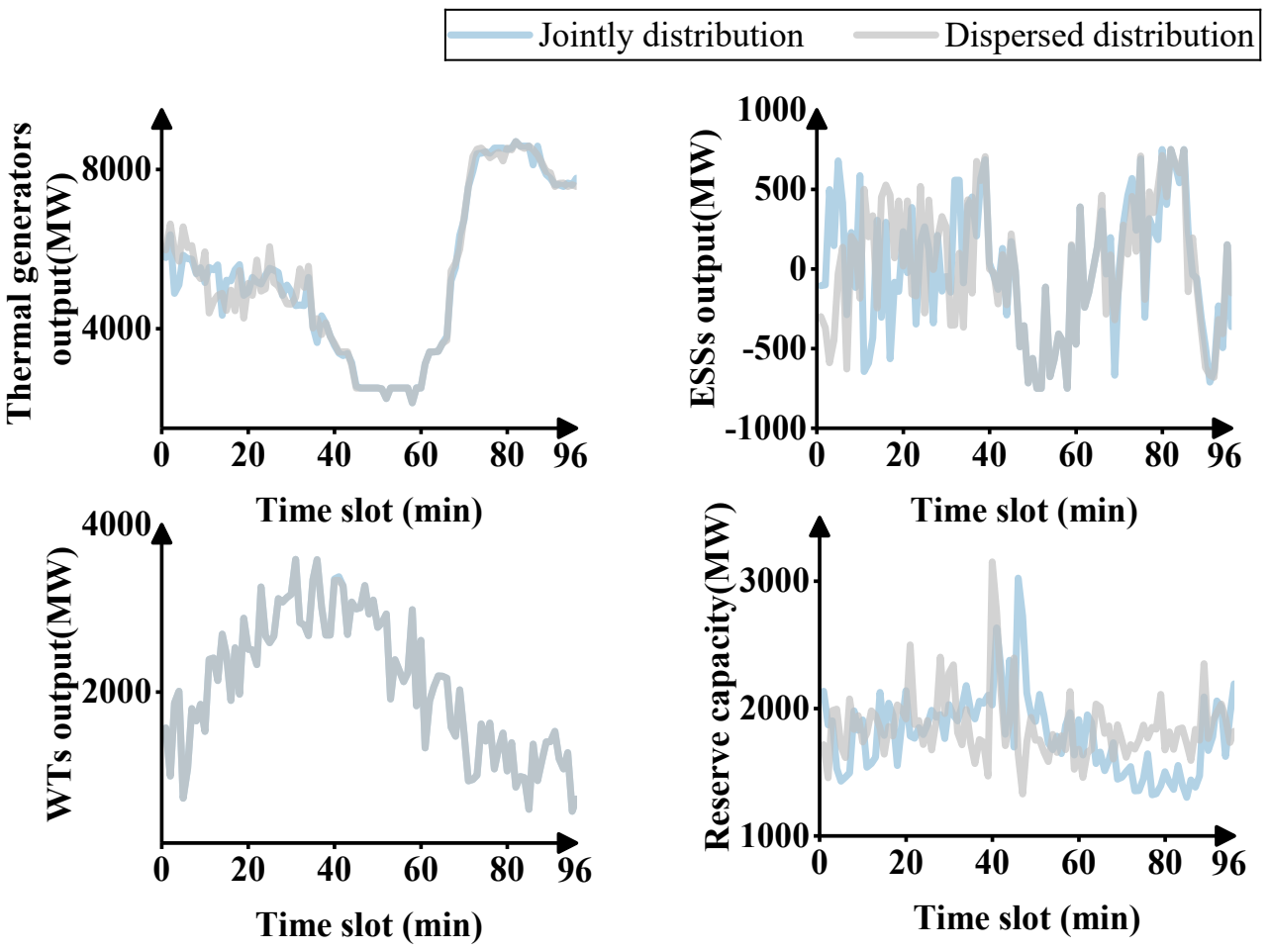

Fig. 13. The day-ahead dispatching schedule of the spatial-joint distribution case and the dispersed distribution case

In terms of frequency indicator, the dispersed distribution method yields a slightly higher frequency nadir margin, with the minimum nadir improved by 0.001 Hz. However, the day-ahead dispatch results show that this method reserves substantially more supporting capacity, with an average increase of 1240.88 MWh over the scheduling horizon. This apparent discrepancy arises because the dispersed distribution approach fails to accurately represent the spatial-joint stochastic behavior of renewable disturbances, leading to an overestimation of uncertainty and a conservative, capacity-driven reserve allocation. A portion of the additional reserves cannot be effectively converted into frequency support due to spatial mismatches and transmission constraints.

2) *Costs improved*

Table II compares the cost of the day-ahead schedules obtained by the two methods. It can be observed that the proposed method achieves lower costs compared with the dispersed-distribution strategy. The generation cost and reserve cost are reduced by 148.36 and 34.14, the unit commitment cost is increased by 4.06 thousand dollars.

The cost reduction stems from mitigating unnecessary unit commitment redundancy enabled by joint-distribution-based disturbance modeling. Treating regional uncertainties as independent tends to overestimate adverse conditions and keeps additional thermal units online as a precaution. By contrast, the proposed method reduces excessive commitments and schedules thermal units closer to cost-efficient operating regions, lowering both commitment and generation costs while satisfying frequency security constraints.

TABLE II
THE COMPARISON OF THE COSTS BETWEEN THE PAPER PROPOSED CASE AND THE DISPERSED DISTRIBUTION CASE

| Case \ Costs( $ ) | Generation costs | Reserve costs | Unit commitment costs | Total costs |
|---|---|---|---|---|
| Joint distribution | $3.42\times10^{6}$ | $1.78\times10^{6}$ | $1.34\times10^{6}$ | $6.53\times10^{6}$ |
| Dispersed distribution | $3.56\times10^{6}$ | $1.81\times10^{6}$ | $1.30\times10^{6}$ | $6.67\times10^{6}$ |

## E. *Sensitivity analysis*

### 1) *Location and magnitude of the contingency*

Sensitivity analysis is conducted to evaluate the impacts of the location and magnitude of the contingency-induced power deficits on regional frequency security. Since the stochastic disturbance setting is determined by the renewable forecast-error distribution and is not directly changed with the contingency location, the forecast error is kept the same as that in the base case. For the contingency-induced power deficits, disturbance locate among three regions are considered. For each location, the contingency magnitude is set to 10%, 15%, and 20% of the maximum load in the corresponding region. Therefore, nine contingency scenarios are constructed, and the corresponding regional nadir and RoCoF results are shown in Fig. 14.

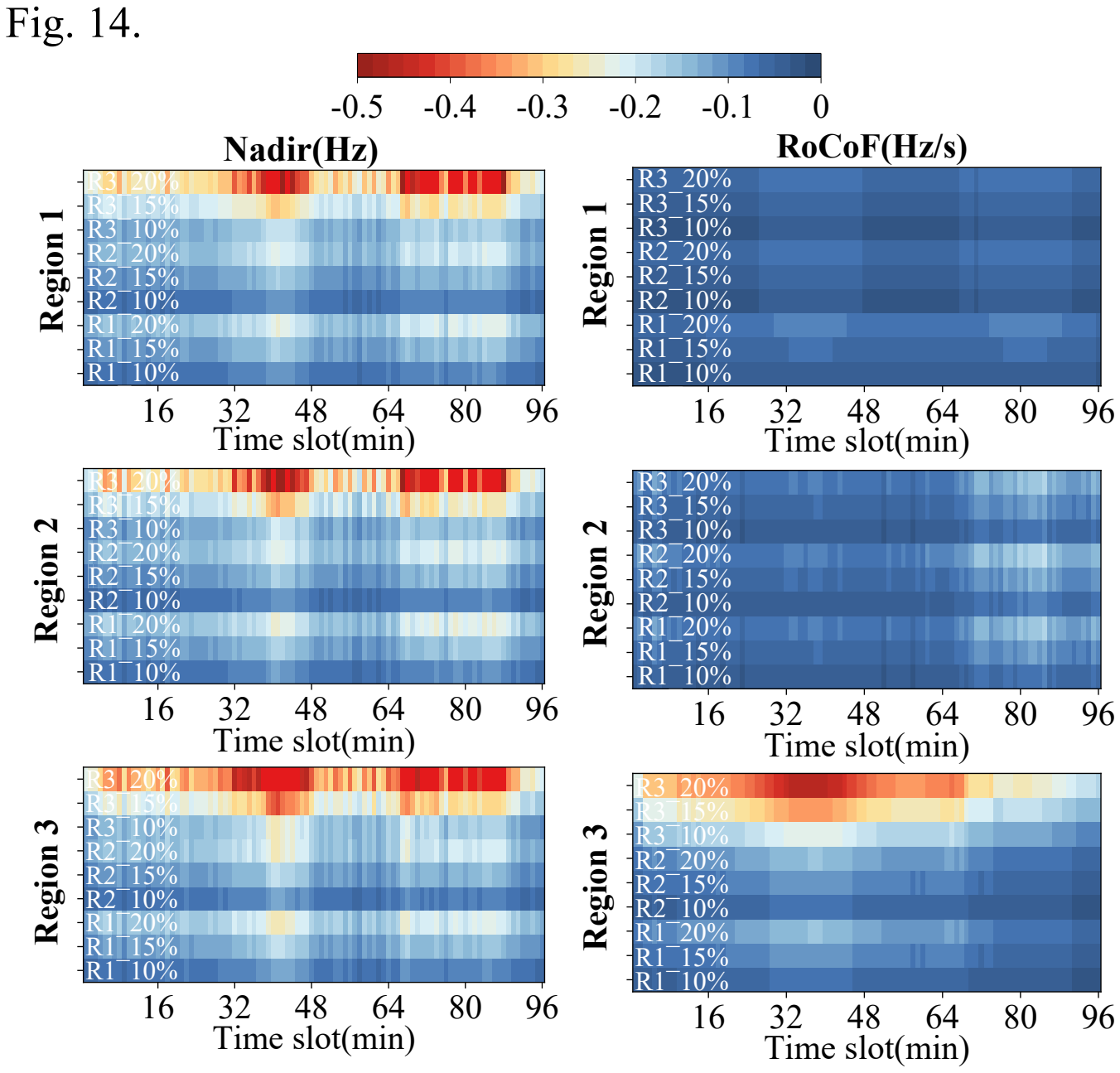

Fig. 14. The nadir and RoCoF indicator under different location and magnitude of the contingency-induced power deficits

As shown in Fig. 14, increasing the contingency magnitude generally leads to more severe nadir values in all regions. From the perspective of disturbance location, the nadir is more significantly affected when the contingency occurs in Region 3. In particular, under the 20% contingency in Region 3, the nadir values of all regions become noticeably lower, indicating that Region 3 is a critical disturbance location for system-wide frequency security. In addition, Region 3 is relatively sensitive to local disturbances, and its RoCoF deteriorates obviously when the contingency magnitude in Region 3 increases. By contrast, when the contingency occurs in Regions 1 or 2, the impact on the overall nadir is less pronounced. For Region 2, the frequency security risk is more strongly affected by renewable forecast-error-induced stochastic disturbances due to its high renewable penetration. Overall, the sensitivity results demonstrate that both the contingency magnitude and disturbance location should be considered in frequency-security-constrained dispatch, and that Region 3 represents the most critical contingency location in the studied system.

### 2) *Limitation of the transmission line capacity*

Sensitivity analysis is conducted to evaluate the impact of inter-regional transmission capacity on system primary frequency reserve allocation. The basic case is defined as the 100% transmission capacity scenario, and the capacity limits of all three inter-regional transmission lines are proportionally scaled to 80%, 90%, 110%, and 120% of the basic case values. Fig. 15 shows the system primary frequency reserve capacities under these three transmission capacity scenarios.

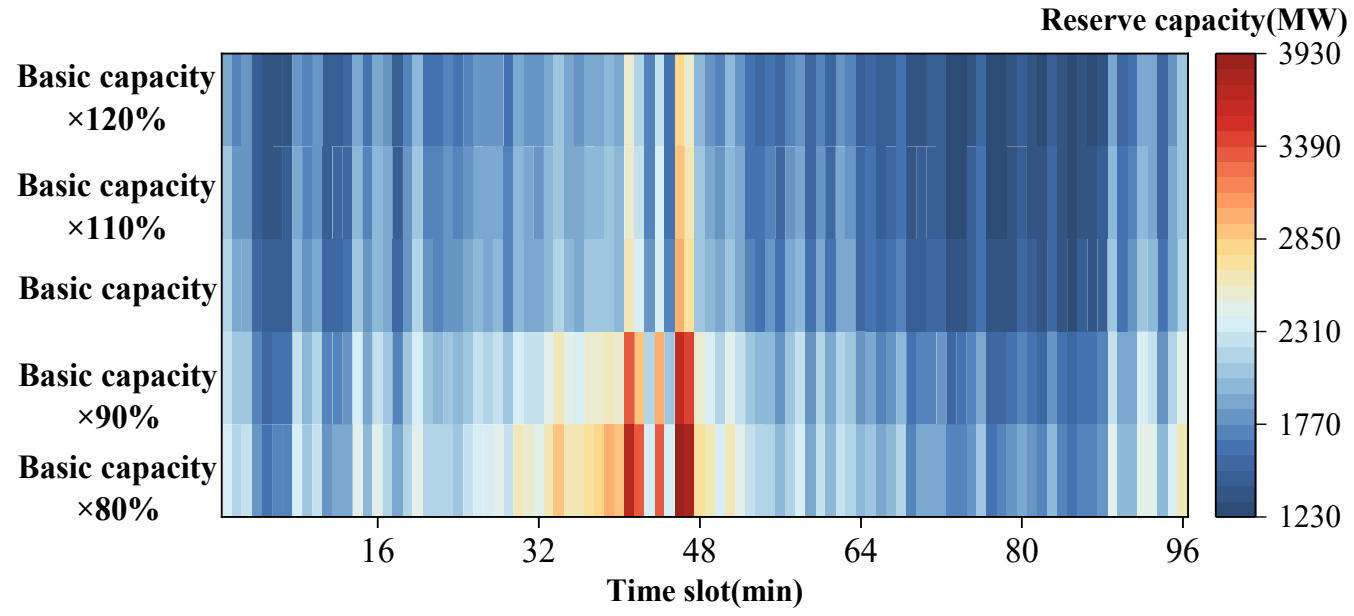

Fig. 15. The system frequency reserve capacities under different transmission line capacity limitation scenarios

Fig. 15 shows that reducing inter-regional transmission capacity weakens cross-regional frequency support and increases the system reserve requirement. In the basic case, the average capacity is 1801.05MW. When the line capacity limits reduce to 80%, the average reserve increased to 2243.29MW, representing a 24.55% increase relative to the basic case reserve capacity. By contrast, when the limits increase to 120%, the average reserve decreased to 1683.94MW, representing a 6.50% reduction relative to the basic case reserve capacity. This asymmetric response occurs because reduced line capacity limits inter-regional reserve deliverability, forcing each region to rely more on local frequency response resources by committing additional thermal generators or deviating from the most economical operating point.

## F. *Analysis of PLB-PINN computing performance*

The comparison is conducted between the proposed PLB-PINN and the numerical Bernstein method [19]. The PLB-PINN is constructed as a multilayer feedforward fully connected neural network with three hidden layers. The numbers of neurons in the three hidden layers are 16, 32, and 16, respectively. ReLU activation functions are adopted in all hidden layers. The PLB-PINN is trained for 500 epochs with a batch size of 32 and a learning rate of $10^{-3}$. The AdamW optimizer is adopted, and the training and testing sets account for 80% and 20% of the dataset, respectively. The numerical Bernstein linearization method approximates the nonlinear nadir expression by first aggregating the system frequency dynamics into a unified COI model and then using multi-segment Bernstein polynomial splines to transform the derivative, integral, and nadir constraints into linear constraints on the spline coefficients. In this study, the frequency response curve is divided into four segments, and a cubic Bernstein polynomial is used in each segment to approximate the COI frequency trajectory, from which the frequency nadir is obtained.

TABLE III
THE COMPARISON OF THE MAE AND ADDITIONAL CONSTRAINTS BETWEEN THE PLB-PINN METHOD AND BERNSTEIN METHOD

| Method | Region | MAE (Hz) | Additional constraints | Computational time (s) |
|---|---|---|---|---|
| PLB-PINN | Reg. 1 | 0.006 | 83520 | 892.58 |
| | Reg. 2 | 0.005 | | |
| | Reg. 3 | 0.022 | | |
| Bernstein method | Reg. 1 | 0.085 | 382656 | 5021.98 |
| | Reg. 2 | 0.062 | | |
| | Reg. 3 | 0.124 | | |

Table III compares the proposed PLB-PINN method with the Bernstein method in terms of mean absolute error (MAE) values, additional constraints, and computational time. As shown in Table III, the proposed PLB-PINN achieves lower MAE values in all three regions. Compared with the Bernstein method, the proposed PLB-PINN reduces the MAE by 92.94%, 91.94%, and 82.26% in Regions 1-3, respectively. In addition, PLB-PINN introduces 83520 additional constraints, while the Bernstein method requires 382656 additional constraints. The number of additional constraints is reduced by 78.17%. The computational time is also decreased from 5021.98s to 892.58s, corresponding to an 82.23% reduction. Therefore, compared with the Bernstein method, the proposed PLB-PINN achieves higher prediction accuracy with a significantly smaller optimization model size and shorter computational time.

The relatively lower fitting accuracy of the Bernstein method mainly arises from two aspects. First, the Bernstein-based formulation is originally developed under a uniform COI frequency response model. When it is applied to the studied multi-region coupled frequency response system, the regional dynamic characteristics are weakened, and the integral term of inter-regional transmission line power exchange is not explicitly represented. Second, the segmented polynomial fitting used in the Bernstein method may introduce discretization and fitting errors, thereby reducing the accuracy of nadir approximation. Meanwhile, from the perspective of optimization complexity, the Bernstein method requires more auxiliary variables and linearized constraints, which enlarges the model size and increases the computational burden. In contrast, the proposed PLB-PINN provides a more accurate and compact representation of the nadir-related frequency-security constraints, thereby achieving a better balance between approximation accuracy and optimization efficiency.

## VII. CONCLUSION

This paper proposes a regional frequency-constrained dispatch method accounting for spatial-joint stochastic disturbances and contingencies, validated on the IEEE 118-bus system. The key findings are as follows:

Firstly, compared with the traditional COI-based method, the proposed method eliminates regional frequency violations masked by system-level indicator. All regional frequency indicator remains within safe thresholds, with the worst-case frequency nadir of vulnerable regions relatively improved by 32.76% and RoCoF optimized by 2.06%. This improvement is achieved with only a 1.31% increase in total dispatch cost, realizing a cost-effective balance between security and economy.

Secondly, the Vine-Copula-based spatial-joint stochastic disturbance modeling avoids the over-conservatism of dispersed distribution methods. It enables precise reserve allocation, reducing total dispatch cost by 2.14% while maintaining frequency security, which verifies the value of accurate uncertainty correlation characterization.

Finally, the PLB-PINN method significantly improves frequency nadir prediction accuracy. Compared with the numerical Bernstein linearization method, the MAE of nadir prediction is relatively reduced by up to 89.05%, and the computational time is reduced by 82.23%, which provides an efficient solution for embedding nonlinear frequency constraints into dispatch optimization.

The proposed method addresses the core challenges of frequency security in renewable-rich multi-regional power systems, offering a practical and scalable optimization framework for engineering applications.